\documentclass[runningheads]{llncs}
\usepackage{graphicx}
\usepackage{amsmath,amssymb}
\DeclareMathOperator{\E}{\mathbb{E}}
\usepackage{multirow}
\begin{document}
\title{A Learning-from-noise Dilated Wide Activation Network for denoising Arterial Spin Labeling (ASL) Perfusion Images}
%
%
\author{Danfeng Xie\inst{1}\and
Yiran Li\inst{1} \and Hanlu Yang\inst{1} \and Li Bai\inst{1} \and Lei Zhang\inst{2} \and Ze Wang\inst{2}
}

\authorrunning{D. Xie et al.}

\institute{Temple University, Philadelphia, PA, 19131, USA \and
University of Maryland School of Medicine, Baltimore, MD, 21201, USA
\email{ze.wang@som.umaryland.edu}}
\maketitle              
\begin{abstract}
Arterial spin labeling (ASL) perfusion MRI provides a non-invasive way to quantify cerebral blood flow (CBF) but it still suffers from a low signal-to-noise-ratio (SNR). Using deep machine learning (DL), several groups have shown encouraging denoising results. Interestingly, the improvement was obtained when the deep neural network was trained using noise-contaminated surrogate reference because of the lack of golden standard high quality ASL CBF images. More strikingly, the output of these DL ASL networks (ASLDN) showed even higher SNR than the surrogate reference. This phenomenon indicates a learning-from-noise capability of deep networks for ASL CBF image denoising, which can be further enhanced by network optimization. In this study, we proposed a new ASLDN to test whether similar or even better ASL CBF image quality can be achieved in the case of highly noisy training reference. Different experiments were performed to validate the learning-from-noise hypothesis. The results showed that the learning-from-noise strategy produced better output quality than ASLDN trained with relatively high SNR reference.

\end{abstract}
\section{Introduction}
Arterial spin labeling (ASL) perfusion MRI provides a non-invasive way to quantify cerebral blood flow (CBF) \cite{detre1992perfusion,williams1992magnetic}. In ASL, the arterial spin labeled image (L image) and the spin untagged image (the control image or C image) were subtracted pair-wisely to generate CBF maps using an appropriate compartment model \cite{alsop2015recommended}. The SNR of ASL CBF maps is inherently low due to the longitudinal relaxation rate (T1) of blood water and the post-labeling transmit process. To improve SNR of the mean CBF maps, a few pairs of L/C images are often acquired. Due to the limitation of total scan time, only 10-50 L/C pairs can be acquired, resulting in a moderate SNR gain by signal averaging. Many post-processing methods have been proposed to denoise ASL MRI \cite{behzadi2007component,wang2012improving,li2018priors} but often with minor to moderate improvement. A main reason is that those methods are based on signal and noise models which are often incomplete or inaccurate for physiological measurements in the case of high noise level.


Nowadays, the major focus of denoising has been increasingly shifted to deep machine learning (DL) given its superb performance for capturing nonlinear and complex data relationship \cite{krizhevsky2012imagenet}. The most widely used deep neural networks consist of multiple layers of receptive field constrained local filters which are trained layer by layer by error backpropagations \cite{krizhevsky2012imagenet} and are often called convolutional neural networks (CNN). The local feature extraction, hierachical abstraction, step-wise backpropagation of CNN and the introduction of several training strategies such as weight drop-out, batch-normalization, skip connection, and residual learning make CNN highly flexible and capable for modeling very complex and nonlinear functions buried in a real-world data such as medical imaging \cite{shen2017deep,litjens2017survey}. 

Several groups have used DL in ASL MRI denoising \cite{kim2017improving,xie2018denoising}. Different from other denoising applications, DL-based ASL denoising network (ASLDN) doesn't have noise-free training references. Accordingly, its denoising performance might be uplimited by the reference image SNR. Interestingly, that potential uplimit doesn't seem to exist as several studies \cite{kim2017improving,ulas2018deepasl,xie2018denoising,xie2019bold,xie2020denoising} showed that ASLDN could produce CBF images with even higher SNR than the reference. Lehtinen et al. \cite{lehtinen2018noise2noise} further investigated this learning-from-noise (LFN) phenomenon in more general settings. The main purpose of this study is to validate that DL-based ASL denoising models can be trained using only noisy image pairs. We show that this new learning-from-noise ASLDN (dubbed as ASLDN-LFN) does not require any quasi-noise-free reference during the training process while it can achieve similar or even better denoising performance than the previous ASLDN that was trained using quasi-noise-free reference images.


\section{Methods}

\subsection{Problem formulation}


Similar to \cite{lehtinen2018noise2noise}, the assumption of ASLDN-LFN is that both the noisy reference $\hat{y}_i$ and the noisy input CBF maps $\hat{x}_i$ are drawn from the same data distribution. When minimizing the L2 loss function: $\frac{1}{N}\sum_i^N (f_{\Theta}(\hat{x}_i)-\hat{y}_i)^2$, a CNN regressor $f_{\Theta}$ is to find the optimum at the arithmetic mean of the observations $ \E\{\hat{y}_i\} $ given enough training samples \cite{lehtinen2018noise2noise}. This training process converges exactly with the process of averaging one subject's all CBF maps to generate quasi-noise-free mean CBF maps (i.e., $ \E\{\hat{y}_i\}) $. Thus, it is not necessary to obtain quasi-noise-free references to train ASLDN.  

However, considering ASL CBFs have excessive outliers, training with L1 loss is preferred. This is because training with L1 loss ($\frac{1}{N}\sum_i |f_{\Theta}(\hat{x}_i)-\hat{y}_i|$) is to find the median of the observations, and the median of the observations is more robust to outliers than the arithmetic mean of the observations \cite{huber1992robust}. We also conducted experiments to compare the effects of training with L1 loss versus training with L2 loss.






\subsection{Network architecture}
 
\begin{figure}
  \centering
  \includegraphics[width=12cm]{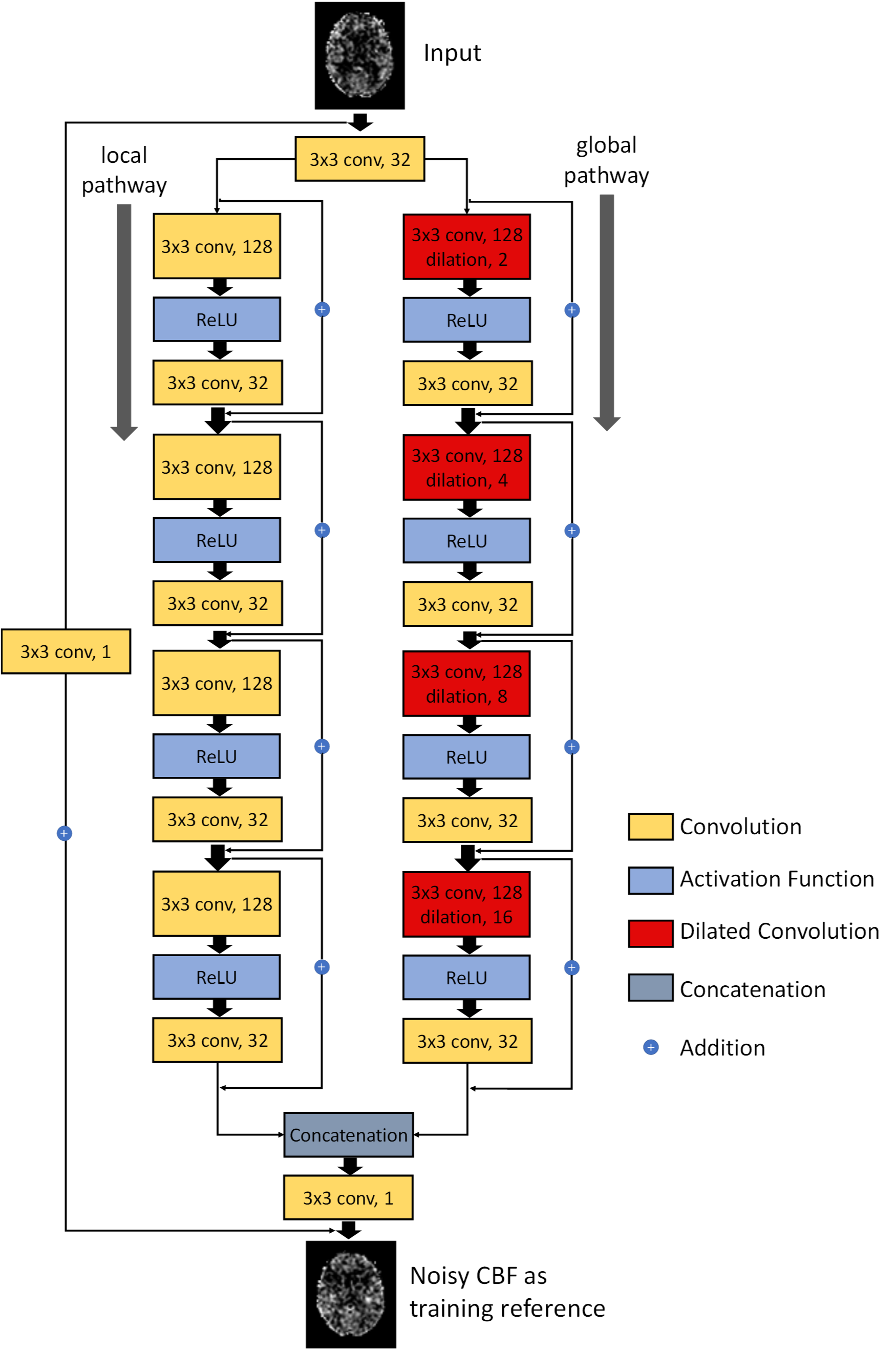}\\
  \caption{Schematic illustration of the architecture of our proposed DWAN network. The first layer consists of 3×3×32 convolutional filters for the input image. Then the output of the first layer was fed to the both local pathway and global pathway. Each pathway contains 4 consecutive wide activation residual blocks. Each wide activation residual block contains two convolutional layers (3×3×128 and 3×3x32) and one activation function layer.  The 3×3×128 convolutional layers in the global pathway were dilated convolutional layers with a dilation rates of 2, 4, 8, 16, respectively. The output of the local pathway and global pathway were concatenated and fed to another 3×3×1 convolutional filter. The 3×3×1 convolutional layer was attached to the end to get the predicted output image with additional input from the input image with 3×3×1 convolution. (a×b×c indicates the property of convolution. a×b is the kernel size of one filter and c is the number of the filters). }\label{network}
\end{figure}

 ASLDN-LFN using the Dilated Wide Activation Network (DWAN) that was proposed in \cite{xie2020denoising}. As shown in Figure \ref{network}, DWAN has two pathways. The difference between the local pathway and global pathway is that the first convolution layer of the 4 wide activation residual blocks in the global pathway used a dilation rate of 2, 4, 8 and 16 respectively. The local pathway extracts the local features and the global pathway uses dilation convolutions to extract global data patterns. Furthermore, the wide activation residual blocks in DWAN are able to expand data features and pass more information through the network, improving performances for low-level computer vision tasks without additional parameters and computation \cite{yu2018wide,fan2018wide}.  By combining the two-pathway structure and the wide activation residual block, this new CNN structure (DWAN) improves the denoising performance in ASLDN-LFN.

\subsection{Data preparation and model training}

 ASL data were pooled from 280 subjects in local database. The data were acquired with a pseudo-continuous ASL sequence (40 control/labeled image pairs with labeling time = 1.5 sec, post-labeling delay = 1.5 sec, FOV=22x22 cm$^2$, matrix=64x64, TR/TE=4000/11 ms, 20 slices with a thickness of 5 mm plus 1 mm gap). ASLtbx \cite{wang2008empirical} was used to preprocess ASL images using the following updated procedures: 1) ASL-specific motion correction method was applied to the raw ASL images (C/L images) to correct systematic label/control labeling induced spurious motions \cite{wang2012improving}; 2) the average of all 40 C/L image pairs was calculated and used as a template for registering the ASL C/L images to the high-resolution T1 image. Registration was performed with SPM12; 3) simple regression was used to regress out residual motions, mean CSF signal, and global signal; 4) adjacent C and L images were subtracted using simple subtraction to generate perfusion-weighted images which were then converted into quantitative CBF using the same method as in \cite{wang2008empirical}.   M0 is approximated by the control image in each label/control image pair and M0 calibration is performed at each voxel separately using the value at the corresponding voxel location of the control image. Outlier CBF image timepoints were identified and removed using the prior-guided slice-wise adaptive outlier cleaning algorithm \cite{li2018priors} ; 5) each subject’s structural MRI was spatially normalized to the Montreal Neurologic Institute (MNI) standard brain using SPM12. The same transform was then applied to the CBF image series.
 
 CBF image slices from 200 subjects were used as the training dataset. CBF images from 20 different subjects were used for validation.  The remaining 60 subjects were used as the testing set.  Input to ASLDN-LFN was the axial slice. All CBF maps were spatially normalized into the Montreal Neurological Institute (MNI) space. For each subject, we extracted one out of every three axial slices from slice 36 to slice 60 of 3D CBF maps in the MNI space and the 2D CBF maps were 109 $\times$ 91 pixels. The 40 ASL CBF images of each subject were divided into 4 time segments, each with 10 successively acquired images. The mean maps of the 1st segment and the 2nd segment were taken as the input and the corresponding reference for DL model training. Another set of input-reference image pairs was obtained from the mean CBF maps of the 3rd and the 4th segment. During model testing, the mean CBF image slices of the first 10 L/C pairs (in the first time segment) were used as the input.

 Due to intrinsic low SNR of ASL MRI, the input and reference CBF maps were already contaminated with severe noise (As Fig. \ref{results}.A. shows). Therefore, no additional artificial noise were added to input and reference CBF maps. Mean CBF maps of the entire 40 L/C image pairs with with Gaussian smoothing (FWHM = 3mm) and state-of-art outlier cleaning \cite{li2018priors} were used as pseudo gold standard $y_i$.  Comparing with previous method ASLDN \cite{xie2018denoising} using pseudo gold standard $y_i$ as training references, the proposed ASLDN-LFN only used noisy data $\hat{y}_i$  as the training reference.   U-Net \cite{xu2017200x} and DilatedNet \cite{kim2017improving}, two popular CNN structure widely used in medical imaging, were implemented as a comparison to our DWAN-based ASLDN-LFN. In all networks, batch-normalization (BN) was removed to avoid the potential errors as demonstrated in \cite{lim2017enhanced} (we noticed that in additional experiments). Additional experiments were conducted to compare the effects of different loss function (L1 and L2) on denoising performance.

 We used Keras and Tensorflow platform to implement all the DL algorithms. Network training was through the adaptive moment estimation (ADAM) algorithm with a learning rate of 0.001 and a batch size of 64. All experiments were performed on a PC with Intel(R) Core(TM) i7-5820k CPU @3.30GHz and a Nvidia GeForce Titan Xp GPU.


\subsection{Evaluation metrics} 

We used Peak signal-to-noise ratio (PSNR) and structure similarity index (SSIM) to quantitatively compare the performance of DWAN with U-Net and DilatedNet. When computing PSNR and SSIM, pseudo gold standard (mean CBF from entire 40 L/C pairs) were used as groundtruth. 

SNR and Grey Matter/White Matter (GM/WM) contrast were calculated to measure the image quality of ASL CBF. The SNR was calculated by using the mean signal of a grey matter (GM) region-of-interest (ROI) divided by the standard deviation of a white matter (WM) ROI in slice 50. The GM/WM contrast was calculated as the mean value of GM masked area divided by the mean value of WM masked area. 

The Correlation coefficient between the DL-produced CBF values and pseudo gold standard were calculated to measure the similarity of the DL-produced CBF values to those processed with non-DL methods. This process was performed at each voxel for ASLDN and ASLDN-FLN separately. The correlation coefficient maps were thresholded by r$>$0.3 for the purpose of comparison and display.


\section{Results}
\begin{figure}[h!]
    \centering
    \includegraphics[width=12cm]{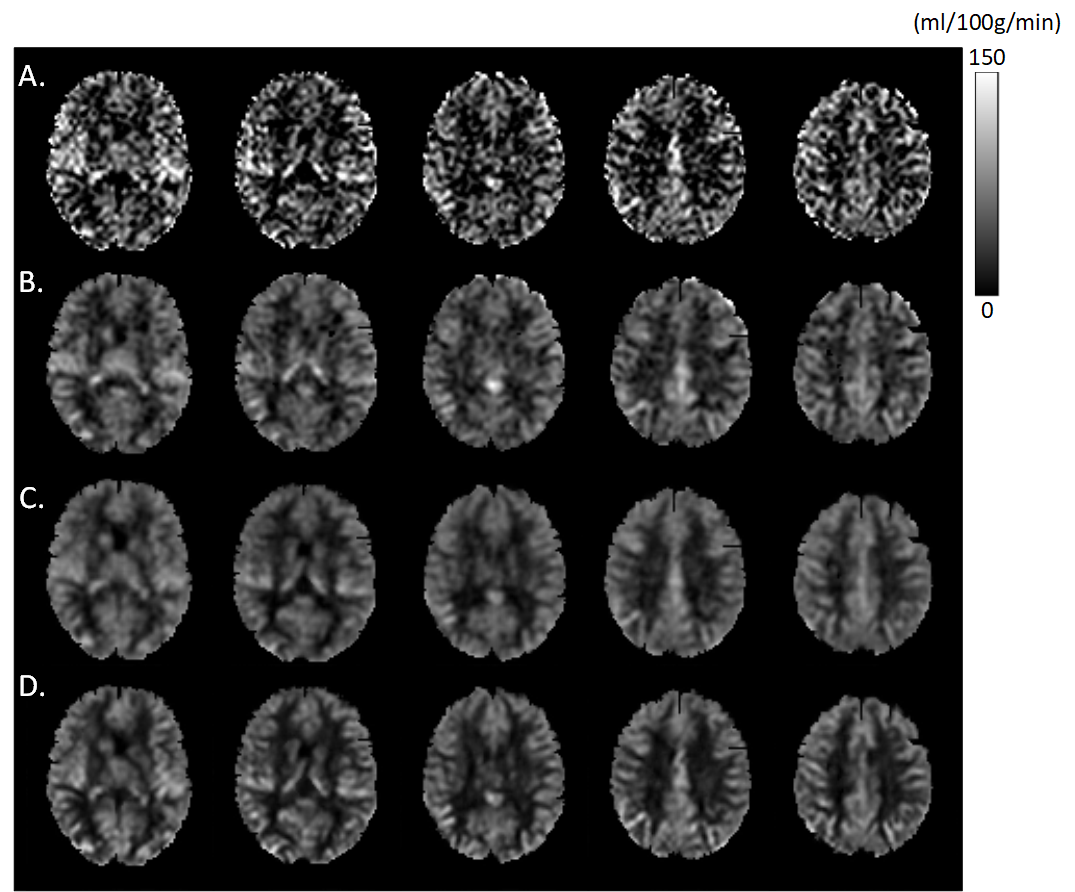}
    \caption{Mean CBF images of a representative subject. The rows from top to bottom are: A. mean CBF maps generated from 10 L/C paris (input to ASLDN-LFN); B. mean CBF maps from all 40 L/C pairs with Gaussian smoothing (FWHM = 3mm) and outlier cleaning (pseudo gold standard); C. output of ASLDN; and D. output of ASLDN-LFN. Only 5 axial slices were shown in each row.}
    \label{results}
\end{figure}

\begin{figure}[h!]
    \centering
    \includegraphics[width=10cm]{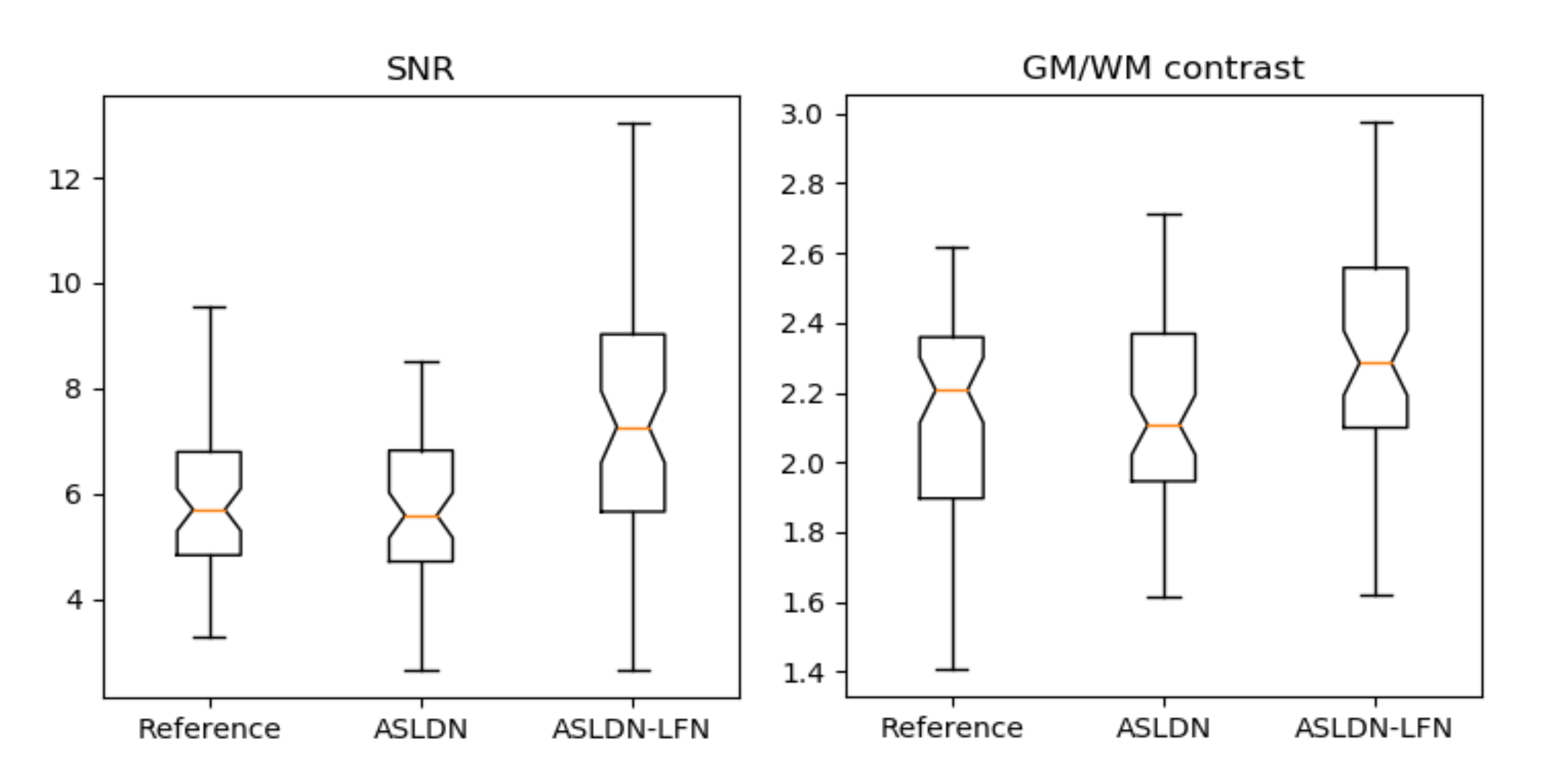}
    \caption{The notched box plot of the SNR (left) and GM/WM contrast (right) from 60 test subjects' CBF maps with different processing methods.}
    \label{SNR}
\end{figure}

Figure \ref{results}  shows the mean CBF maps produced by different algorithms.  Compared to pseudo gold standard (Fig. \ref{results}.B.) and the output of ASLDN (Fig. \ref{results}.C.), the CBF maps produced by ASLDN-LFN (Fig. \ref{results}.D.) showed substantially improved quality in terms of suppressed noise and better perfusion contrast between tissues. Figure \ref{SNR} shows the notched box plot of the SNR and GM/WM contrast from 60 test subjects' mean CBF maps processed with different methods. The average SNR of pseudo gold standard, the output of ASLDN and the output of ASLDN-LFN were 5.87, 6.36 and 8.06 respectively. The average GM/WM contrast of pseudo gold standard, the output of ASLDN and the output of ASLDN-LFN were 2.14, 2.15 and 2.32. ASLDN-LFN achieved better SNR and GM/WM contrast than ASLDN and pseudo gold standard (paired t-test, $p<0.05$).

\begin{figure}[h!]
    \centering
    \includegraphics[width=10cm]{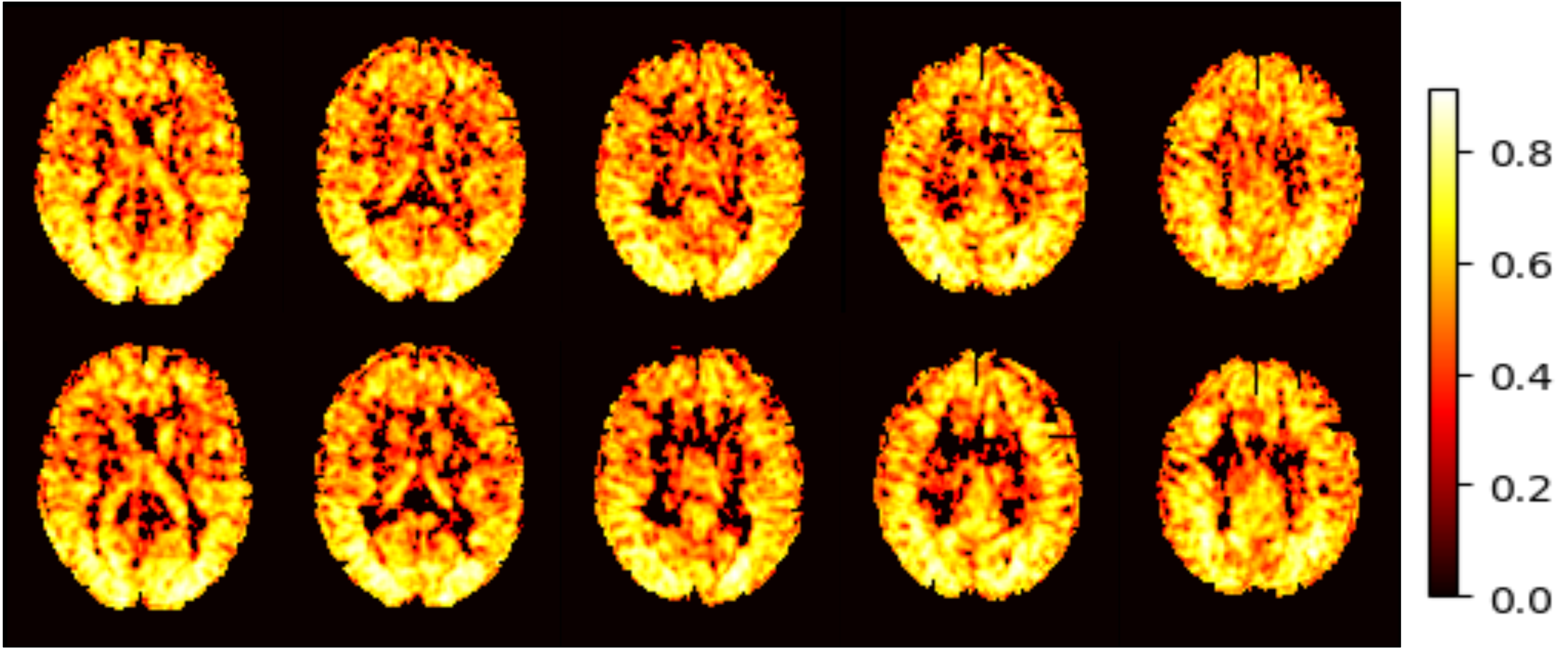}
    \caption{Correlation coefficient maps of ASLDN (top) and ASLDN-LFN (bottom). Only 5 axial slices were shown. Correlation coefficients less than 0.3 were thresholded to be 0.}
    \label{GMWM}
\end{figure}

Figure \ref{GMWM} shows the correlation coefficient maps of ASLDN and ASLDN-LFN. Correlation coefficient at each voxel was calculated between the pseudo gold standard and network output. Outputs of ASLDN and ASLDN-LFN strongly correlated to the pseudo gold standard, proving that both networks can preserve individual subjects' CBF patterns while suppressing noise. Output of ASLDN-LFN showed less correlation to input in WM because ASLDN-LFN removed more noises in WM than ASLDN.

\begin{table}[h!]
    \centering
    \begin{tabular}{|c|c|c|c|c|c|c|}
 \hline    & \multicolumn{3}{|c|}{ ASLDN} & \multicolumn{3}{|c|}{ ASLDN Learning-from-noise} \\
    \hline 
       Model& U-Net   & DilatedNet & DWAN  & U-Net & DilatedNet & DWAN  \\ \hline
         PSNR  & 24.53& 24.92& \textbf{25.26} & 24.84 & 25.06 & \textbf{25.28}\\\hline
         SSIM & 0.796 & 0.793& \textbf{0.802}& 0.798 & 0.797  &\textbf{0.803}\\ \hline
    \end{tabular}
    \caption{The average PSNR and SSIM of mean CBF maps produced by different CNN architectures in different training schemes.}
    \label{models}
\end{table}

Table \ref{models} lists the PSNR and SSIM performance of ASLDN \cite{xie2018denoising} and the proposed ASLDN-LFN with or without using the DWAN network structure. ASLDN-LFN showed higher PSNR and SSIM than previous ASLDN. Using DWAN in ASLDN and ASLDN-FLN provided higher PSNR and SSIM than U-Net and DilatedNet.

\begin{figure}[h!]
    \centering
    \includegraphics[width=12cm]{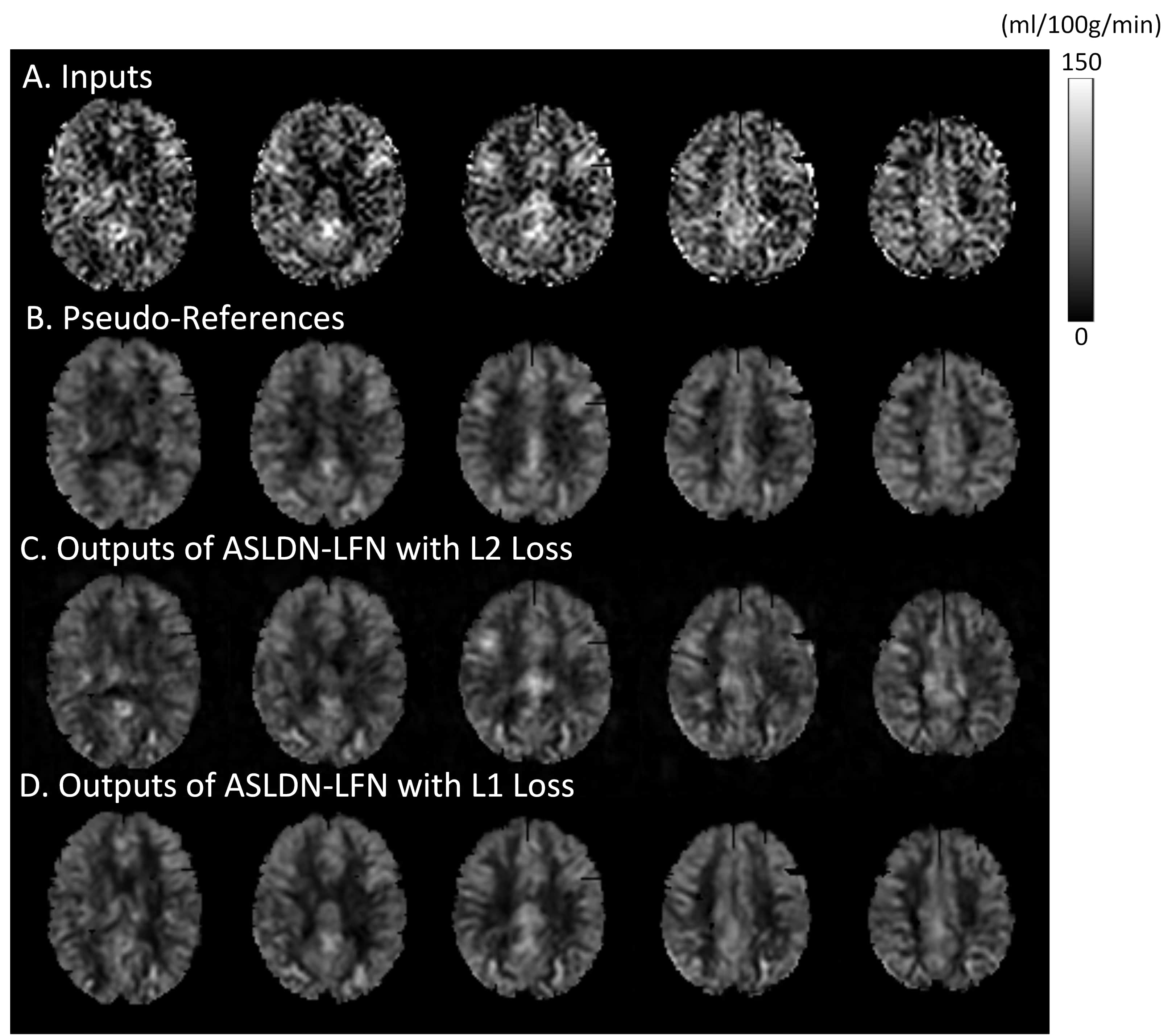}
    \caption{Mean CBF maps from one representative subject (Only 5 axial slices were shown in each row). The rows from top to bottom are: A. mean CBF maps generated from 10 L/C paris (input to ASLDN-LFN); B. mean CBF maps from all 40 L/C pairs with Gaussian smoothing (FWHM = 3mm) and outlier cleaning (pseudo gold standard); C. outputs of ASLDN-LFN trained with L2 loss; D. and outputs of ASLDN-LFN trained with L1 loss }
    \label{L1L2}
\end{figure}

Figure \ref{L1L2} shows the result of ASLDN-LFN that was trained with different loss functions. When input mean CBF maps contained large amounts of outliers, ASLDN-LFN trained with L2 loss was affected, resulting in deteriorated perfusion in grey matter area. ASLDN-LFN trained with L1 loss, in contrast, remains unaffected due to its robustness to outliers. PSNR and SSIM are used to quantitatively measure the denoising performance of ASLDN-LFN trained with L1 loss and L2 loss. PSNR and SSIM were 24.40 and 0.677 when ASLDN-LFN was trained with L2 loss, whereas PSNR and SSIM were 25.28 and 0.803 When ASLDN-LFN was trained with L1 loss.

\section{Conclusion}
In this study, we propose ASLDN-LFN to show that DL-based ASL denoising models can be trained using only noisy image pairs. The experimental results demonstrate that ASLDN-LFN can reliably denoise ASL CBF and even achieve improved image quality than ASLDN in terms of SNR and GM/WM contrast. Besides, we show that training using L1 loss is more robust to outliers than training using L2 loss for ASLDN-LFN.  Moreover, by using ASLDN-LFN, more training data can be generated as it requires less L/C pairs to generate reference mean CBF maps, which is particular useful when ASL CBF data are limited.

\section*{Acknowledgement}
This work was supported by NIH/NIA grant: 1 R01 AG060054-01A1
%
%

%
%
%
\bibliographystyle{splncs04}
\bibliography{main}
\end{document}